\documentclass[12pt,preprint]{aastex}
\newcommand{\kms}{\,{\rm km\,s^{-1}}}
\newcommand{\msun}{\,{\rm M_\odot}}

\newcommand{\etal}{{et al.\ }}
\newcommand{\vir}{{\rm vir}}
\newcommand{\gas}{{\rm gas}}

\newcommand{\beq}{\begin{equation}}
\newcommand{\eeq}{\end{equation}}
\newcommand{\ba}{\begin{eqnarray}}
\newcommand{\ea}{\end{eqnarray}}
\def\spose#1{\hbox to 0pt{#1\hss}}
\newcommand{\lta}{\mathrel{\spose{\lower 3pt\hbox{$\mathchar"218$}}
      \raise 2.0pt\hbox{$\mathchar"13C$}}}
\newcommand{\gta}{\mathrel{\spose{\lower 3pt\hbox{$\mathchar"218$}}
      \raise 2.0pt\hbox{$\mathchar"13E$}}}
\shortauthors{Volonteri \& Rees}
\shorttitle{Rapid growth of high redshift black holes}

\begin{document}
\title{Rapid growth of high redshift black holes}
\author{Marta Volonteri\altaffilmark{1}, Martin J. Rees\altaffilmark{1}}
\altaffiltext{1}{Institute of Astronomy, Madingley Road, Cambridge
CB3 0HA, UK}

\begin{abstract}
We discuss a model for the early assembly of supermassive black holes (SMBHs)
at the center of galaxies that trace their hierarchical build-up far up in the
dark halo `merger tree'. Motivated by the observations of luminous quasars 
around redshift $z\approx 6$ with SMBH masses $\approx 10^9\msun$, we assess the 
possibility of an early phase of stable super-critical quasi-spherical accretion in the
BHs hosted by metal free halos with virial temperature $T_{\rm vir} > 10^4$K. 
We assume that the first `seed' black holes formed with intermediate masses 
following the collapse of the first generation of stars,
in mini-halos collapsing at $z\sim 20$ from high-$\sigma$ density fluctuations. 
In high redshift halos with $T_{\rm vir} > 10^4$K, conditions exist for 
the formation of a fat disc of gas at $T_\gas\approx 5000-10000$K. Cooling via 
hydrogen atomic lines is in fact effective in these comparatively massive halos.
The cooling and collapse of an initially spherical configuration of gas leads to
a rotationally supported disc at the center of the halo if baryons preserve their 
specific angular momentum during collapse. The conditions for the formation of
the gas disc and accretion onto a central black holes out of this supply of gas are
investigated, as well as the feedback of the emission onto the host and onto
the intergalactic medium. We find that even a short phase of super-critical accretion
eases the requirements set by the $z\approx 6$ quasars.

\end{abstract}

\section{Introduction}
The strongest constraint on the high-redshift evolution of supermassive black holes (SMBHs)
comes from the observation of luminous quasars at $z\approx 6$ in the Sloan 
Digital Sky Survey (SDSS, Fan \etal 2001). The luminosities of these quasars, well in excess 
of $10^{47}$ erg/s imply that SMBHs with masses $\approx 10^9 M_\odot$ are already in place 
when the Universe is only {\rm 1 Gyr} old. 
The highest redshift quasar currently known, SDSS 1148+3251, at $z=6.4$,
has estimates of the SMBH mass in the range $(2-6)\times 10^9 M_\odot$ (Barth et al. 2003, 
Willott, McLure \& Jarvis 2003).  Among the seed BHs proposed, the less exotic 
(e.g. PopIII star remnants, gravitationally collapsed star clusters) are in the range 
$10^2-10^4 M_\odot$, forming at $z=30$ or less. 

To grow such seeds up to $10^9 M_\odot$ 
requires an almost continuous accretion of gas, assuming Eddington accretion and 
a standard thin disc radiative efficiency for a Schwarzschild black hole, $\epsilon\simeq0.1$. 
If accretion is via a gaseous thin disc, though, the alignment of a SMBH with the angular 
momentum of the accretion disc tends to efficiently spin holes up (Volonteri et al 2004), 
and radiative efficiencies can therefore approach 30\% (assuming a ``standard'' 
spin-efficiency conversion). The accretion of mass at the Eddington rate causes the BH
mass to increase in time as
\beq
M(t)=M(0)\,\exp\left(\frac{1-\epsilon}{\epsilon}\frac{t}{t_{\rm Edd}}\right),
\eeq
where $t_{\rm Edd}=0.45\,{\rm Gyr} $. Given $M(0)$, the higher the 
efficiency, the longer it takes for the BH to grow in mass by (say) 20 e-foldings (Shapiro 2005).

Yu \& Tremaine (2002), Elvis, Risaliti, \& Zamorani (2002), and Marconi
\etal (2004), compared the local MBH mass density to the mass density accreted by luminous 
quasars, showing that quasars have a mass-to-energy conversion efficiency $\epsilon\gta 0.1$ 
(an argument originally proposed by Soltan 1982).  This high average radiative 
efficiency, though, describes the population of SMBHs at redshift $z<5$, and nothing is known
about the radiative efficiency of pregalactic quasars in the early Universe. 
The numbers quoted above, in fact, may suggest that the picture that we have of the low 
redshift Universe may not apply at earlier times.

In this paper we follow earlier papers in considering a scenario for the hierarchical 
assembly of SMBHs that traces their seeds back to the very first generation of stars,
in mini-halos above the cosmological Jeans mass collapsing at $z\sim 20$ from the
high-$\sigma$ peaks of the primordial density field. 

However we introduce two new features: (i) we consider more explicitly the configuration of the 
gas from which the accretion occurs:  a fat, dense disc of cold gas, likely to
form in halos with $T_{\rm vir}>10^4 {\rm K}$ and zero metallicity (Oh \& Haiman 2002).
And (ii) we explore the assumption that at early stages, accretion occurs at a 'Bondi' rate 
that is higher than the standard rate for $\epsilon\simeq0.1$

The evolution of the \emph{first} MBHs into the \emph{first}
pregalactic quasars and their impact on the re-ionization of the universe in this scenario
has been explored recently by Madau \etal 2004. We here expand and deepen the previous study, 
focusing in particular on the interplay of gas cooling and BH feeding. 

In the next section we review the model for assembly of MBHs in cold dark matter (CDM)
cosmogonies.  We then address the conditions of the gas around the MBHs in the cores of 
high redshift mini-halos (\S3), before and after the onset of pregalactic quasar activity. 
We compute the global evolution of the MBH population and discuss its implications in \S4 and finally 
summarize our results in \S5. Unless otherwise stated, all results shown below refer to the 
currently favoured $\Lambda$CDM world model with $\Omega_M=0.3$, $\Omega_\Lambda=0.7$, 
$h=0.7$, $\Omega_b=0.045$, $\sigma_8=0.93$, and $n=1$.

\section{Assembly of pregalactic MBH\lowercase{s}}

The main features of a plausible scenario for the hierarchical
assembly, growth, and dynamics of MBHs in a $\Lambda$CDM cosmology
have been discussed by Volonteri, Haardt, \& Madau (2003), Volonteri 
\etal (2004), Madau \etal (2004).   ``Seed" holes with intermediate masses
form as end-product of the very first generation of stars. 
They form in isolation within mini-halos above the cosmological Jeans mass 
collapsing at $z\approx24$ from rare $\nu$-$\sigma$ peaks of the primordial density 
field (Madau \& Rees 2001). Pregalactic seed IMBHs form within the mass ranges 
$20<m_\bullet<70\,\msun$ and $130<m_\bullet<600\,\msun$ (Fryer, Woosley, \& Heger 2002, 
Omukai \& Palla 2003); for simplicity, 
within these intervals the differential black hole mass function is assumed to be flat. 
As our fiducial model we take $\nu=4$, corresponding to a mass density parameter in
MBHs of $\Omega_\bullet\sim10^{-9}$. 
This is much less than the mass density of the supermassive variety 
found in the nuclei of most nearby galaxies, $\Omega_{\rm SMBH}=(2.1\pm
0.3) \times 10^{-6}$ (Yu \& Tremaine 2002).  
Note that this choice of seed MBHs occupation is similar to that 
of Volonteri et al. 2003 (seed holes in $3.5\sigma$ peaks at $z=20$).

\section{Super-critical accretion and rapid growth of SMBHs}
\subsection{Fat cold discs in high redshift haloes}
Metal--free gas in halos with virial temperatures $T_{\rm vir} > 10^4$K can cool in
the absence of ${\rm H_2}$ via hydrogen atomic lines to $\sim 8000$ K. 
Oh \& Haiman (2002) considered  the cooling and collapse of an initially
spherical configuration of gas in a typical halo with $T_{\rm vir} >
10^4$K. In the absence of ${\rm H_2}$ molecules, the gas
can contract nearly isothermally at this temperature. Since cooling via the
Ly$\alpha$ line has a very sharp cutoff at $T < 10^4$K (Spitzer
1978) a 'thermostatic mechanism' operates: if the gas cools below $\sim 10^4$K, 
the gas recombines and the cooling time rapidly increases. The gas then heats 
and contract until atoms are collisionally re--ionized.  

Halos, though, possess angular momentum, $J$, that can be related to the spin
parameter $\lambda \equiv J |E|^{1/2}/G M_h^{5/2}$, where $E$ and $M_h$ are the 
total energy and mass of the halo. The baryons probably preserve their specific angular
momentum during collapse (Mo, Mao \& White 1998), and settle into a rotationally supported 
disc at the center of the halo (Mo, Mao \& White 1998, Oh \& Haiman 2002).

Following Oh \& Haiman (2002), we assume that a fraction $f_d$ of
the gas settles into an isothermal, exponential disc, with gas temperature 
$T_{\rm gas}$, embedded in a dark matter halo of virial temperature $T_{\rm vir}$, described by a
Navarro, Frenk \& White (1997, hereafter NFW) density
profile.  The mass of the disc can therefore be expressed as
$M_{\rm disc}= f_d (\Omega_b/\Omega_M) M_h$. 

Under these assumptions, the disc scale radius is given by 
$R_d \sim 2^{-1/2} \lambda R_{\rm vir}$. 

For an isothermal exponential disc of radial scale length $R_d$, 
the number density of hydrogen at radius $r$ and at vertical height $z$ can be 
written as 
\begin{equation}
n(r,z)=
 n_0 
 {\rm exp}\left(-\frac{2 r}{R_d} \right) 
 {\rm sech^{2}} \left(\frac{z}{\sqrt{2} z_0} \right),
\label{disc_density}
\end{equation}
where $n_0$ is the central density and $z_0$ is the vertical scale height
of the disc at radius $r$.

The central number density of the gas can be written as:
\begin{eqnarray}
n_0 &\simeq& 6 \times 10^{4} 
f_{d,0.5}^{2} \,
\lambda_{0.05}^{-4} \,
T_{\rm gas,8000}^{-1}\,
R_{\rm vir,6}^{-4} \,
M_{h,9}^{2} \, {\rm cm^{-3}}.
\label{central_density}
\end{eqnarray}
Here $M_{h,9}$ is the halo mass in units of $10^9\msun$, $T_{\rm gas,8000}$ is the gas 
temperature in units of 8000K, $R_{\rm vir,6}$  is the virial radius in units of 6kpc, 
$f_{d,0.5}$ is the fraction of gas in the disc normalized to $0.5$,
and $\lambda_{0.05}$ the spin parameter normalized to 0.05. 
Given the scaling of halo mass and virial radius in the spherical collapse model, 
$n_0\propto \lambda^{-4} f_d^2(T_{\rm gas}/T_{\rm vir})^{-1}$, so
$z_0\propto\lambda^2 f_d^{-1}(T_{\rm gas}/T_{\rm vir})^{1/2}$ and the disc is 'fat' 
when $T_{\rm gas}\approx T_{\rm vir}$, i.e. for haloes with virial temperature
$T_{\rm vir} \gta 10^4$K whose gas can cool to $\sim 8000$ K due to hydrogen atomic line
cooling. The disc thins out if the gas cools to lower temperatures in presence of metals 
or ${\rm H_2}$ molecules.

Oh \& Haiman showed that if only atomic line cooling operates, the discs are stable 
to fragmentation in the majority of cases, and this fat disc contains essentially all 
the baryons. In the rare cases  where
fragmentation is possible, there would presumably then be star formation. 
However it seems inevitable that a substantial fraction of the baryons 
will remain in the fat disc at $5000{\rm K}<T_\gas<10^4{\rm K}$
(at the bottom of this range if there were no significant heat input, but shifting 
towards the peak of the HI cooling curve if there is substantial heating).
\subsection{Accretion onto a central MBH}
As described in \S2, we assume that one seed MBH forms in each of the rare density
peaks above 4-$\sigma$ at $z=24$. The MBH is supposed to be at rest in the center of the 
halo, after the formation and collapse of a PopIII star in the central region of the mini-halos.

Let us now consider a MBH seed in a metal free halo with $T_\vir\gta 10^4\,$K. 
As delineated in the previous section, a fat disc of cold 
($5000{\rm K}<T_\gas<10^4{\rm K}$) gas likely condenses at the center of the halo. 
This gas can supply fuel for accretion onto a MBH within it.  
We will now discuss how stable super-critical accretion  (i.e.
much larger than the Eddington rate) of the gas can proceed.

In a spherical geometry, matter can infall onto the hole at a rate that can be 
estimated using the Bondi-Hoyle formula (Bondi \& Hoyle 1944):
\beq
\dot{M}_{\rm Bondi}=\frac{\alpha \,4\pi\, G^2\,M_{\rm BH}^2\, m_{\rm H} n}{c_s^3}
\label{eq:dotm}
\eeq
where $\alpha$ is a dimensionless parameter of order unity and $c_s \sim 10(T/10^4{\rm K})^{1/2}\,\kms$ 
is the sound speed of the gas. 

It is useful to compare the above accretion rate to the Eddington rate:
\beq
\frac{\dot{M}_{\rm Bondi}}{\dot{M}_{\rm Edd}}=
40\,M_{\rm BH,3}\,n_{0,4}\,T_{\rm gas,8000}^{-3/2},
\label{eq:dotmEdd}
\eeq
where the central density $n_{0,4}$ is in units of $10^4 {\rm cm^{-3}}$ and the MBH mass, $M_{\rm BH,3}$,
is normalized to $10^3\msun$. Clearly for the typical black hole masses and gas densities we are 
considering here, $\dot{M}_{\rm Bondi}>>\dot{M}_{\rm Edd}$. 

The Bondi radius for the  MBH is  
$r_{\rm acc}=G\, M_{\rm BH}/c_s^2\simeq0.05\, {\rm pc} (M_{\rm BH}/10^3\msun)(T_\gas/8000{\rm K})^{-1}$.
Let us now compare the Bondi accretion radius to the fat disc thickness:

\begin{eqnarray}
\nonumber
\frac{r_{\rm acc}}{z_0}&\sim&\frac{G\,M_{\rm BH}}{c_s^3} 
(4 \pi G\,\mu m_{\rm H} n_0)^{1/2}=\\
&&
\nonumber
\hspace{-0.3cm}
6\times 10^{-2} \,M_{\rm BH,3}\,
T_{\rm gas,8000}^{-1.5}\,
n_{0,4}^{1/2}\,\lambda_{0.05}^{-2}.
\label{rb_z0}
\end{eqnarray}.

For MBHs in halos with $T_\vir\gta 10^4\,$K, the accretion radius is then
well within the disc, surrounded by gas with almost constant density
$n_0$ and thus a quasi-spherical geometry. 
In particular, if the gas disc rotates as a rigid body, the transverse velocity at the Bondi 
radius is very small compared to the free fall velocity. The latter is, by definition, equal
to the sound speed, while the former is a factor ${r_{\rm acc}}/R_d$ smaller than $v_D$, the 
rotation velocity of the gas disc at $R_d$, $v_D \gtrsim c_s$. The angular momentum is nonetheless 
too large for the gas to fall radially into the hole and a tiny accretion disc forms. 
The outer edge of the accretion disc, $r_{\rm in}$, can be expressed as:
\beq
\frac{r_{\rm in}}{r_S}=3\times10^2\,v_{D,10}^2\,M_{\rm BH,3}^2\,\lambda_{0.05}^{-2}\,R_{\rm vir,6}^{-2}
\label{r_inner}
\eeq
where $r_S$ is the Schwarzschild radius of the hole, $v_D$ is in units of $10\,\kms$ and 
we have assumed specific angular momentum conservation inside $r_{\rm acc}$.
The accretion disc has therefore a size of order of the trapping radius: 

\beq
r_{\rm tr}=r_S\frac{\dot M}{\dot{M}_{\rm Edd}}=0.12\,r_{\rm in}\,v_{D,10}^{-2}\,M_{\rm BH,3}^{-1}\,\lambda_{0.05}^2
\,R_{\rm vir,6}^2n_{0,4}, 
\label{r_tr}
\eeq
i.e. the radius
at which radiation is trapped as the infall speed of the gas is larger than the 
diffusion speed of the radiation. 
The trapping radius is equivalent to the radius of spherization defined by
Shakura \& Sunyaev (1973), where the thickness of the disc becomes of the same order 
as $r_{\rm tr}$; at smaller radii the the inflow is quasi-spherical. 

Begelman (1979) and Begelman \& Meier (1982) studied super-critical accretion onto a 
BH in spherical geometry and quiescent thick discs respectively. In the spherical
case, radiation pressure cannot prevent the accretion rate from being super-critical, 
while the emergent luminosity is limited to $\lta L_{\rm Edd}$. 
The case with angular momentum is more complicated (especially with regard to the role of winds).
Though this issue remains unclear, it still seems  possible that when the inflow rate is 
super-critical, the radiative efficiency drops so that the hole can accept the material without 
greatly exceeding the Eddington luminosity.  The  efficiency could be low either because most 
radiation is trapped and advected inward, or because the flow adjusts so that the material can 
plunge in from an orbit with small binding energy (Abramowicz \& Lasota 1980).

Despite the uncertainties it seems worthwhile to explore the consequences for black hole growth of 
an early phase when the MBH can accept most the mass infalling at a rate that can be estimated 
using the Bondi-Hoyle formula, and can therefore grow much more rapidly than the Eddington rate 
would allow.

The conditions for quasi-spherical geometry hold when the accretion radius is much smaller 
than the disc thickness, i.e. when the MBH and the halo are small and the gas temperature 
is $\sim 8000$ K. The gas temperature will not be much larger than $\sim 10^4$ K, as the dense gas 
can radiate away all the energy injected by the quasar in photoionizing photons.

On the other hand, if the gas is able to cool down to $10{\rm K}<T_\gas<200{\rm K}$, due to formation of 
molecular hydrogen or metal pollution, collapse and fragmentation in the disc
will suddenly halt the Bondi-style accretion. We therefore argue that this super-critical accretion
phase would end when the Universe is enriched by metals at $6<z<10$.

In the absence of metals, ${\rm H_2}$ would be the main coolant, but 
the strong internal feedback provided by the  pregalactic quasars emission
within a galaxy easily suppresses ${\rm H_{2}}$ formation. 
In fact, assuming a nonthermal power-law component ($\alpha=1$), in the 11.15-13.6 eV range,
the total rate of dissociations produced by the pregalactic quasars in the gas disc is 
$\dot{N}_{\rm diss}\sim 10^{53}\, x_{\rm H_2} \,(M_{\rm BH}/10^3\msun)
(n_0/10^4{\rm cm^{-3}})(z_0/pc) \,{\rm s^{-1}}$. The total number of
molecules in the disc is $N_{\rm H_2}\sim 10^{53}\,x_{\rm H_2} (M_{\rm disc}/10^6\msun)$.
The timescale for dissociating all the ${\rm H_2}$ molecules present in the disc
is thus very short. While the pregalactic quasar is shining, ${\rm H_2}$ formation 
is therefore completely halted in the disc, and cooling below $\sim 4000$K is impossible.

\section{Results}
To assess the relative importance of this processes, we trace the evolution
of a MBH along with its host halo. We generate Monte Carlo realizations (based on 
the extended Press-Schechter formalism) of the merger hierarchy of a $M_h=10^{13}\msun$ 
halo at $z=6$. The halo mass is chosen by requiring that the number density in haloes 
more massive than $M_h$ matches the space density of quasars at $z=6$ (Fan et al. 2003,
2004). We then extract from the trees the mass-growth history of the main halo and of 
smaller satellites (selecting only density peaks above 4-$\sigma$) and grant these halos 
a MBH seed. 

\begin{figure}
\plotone{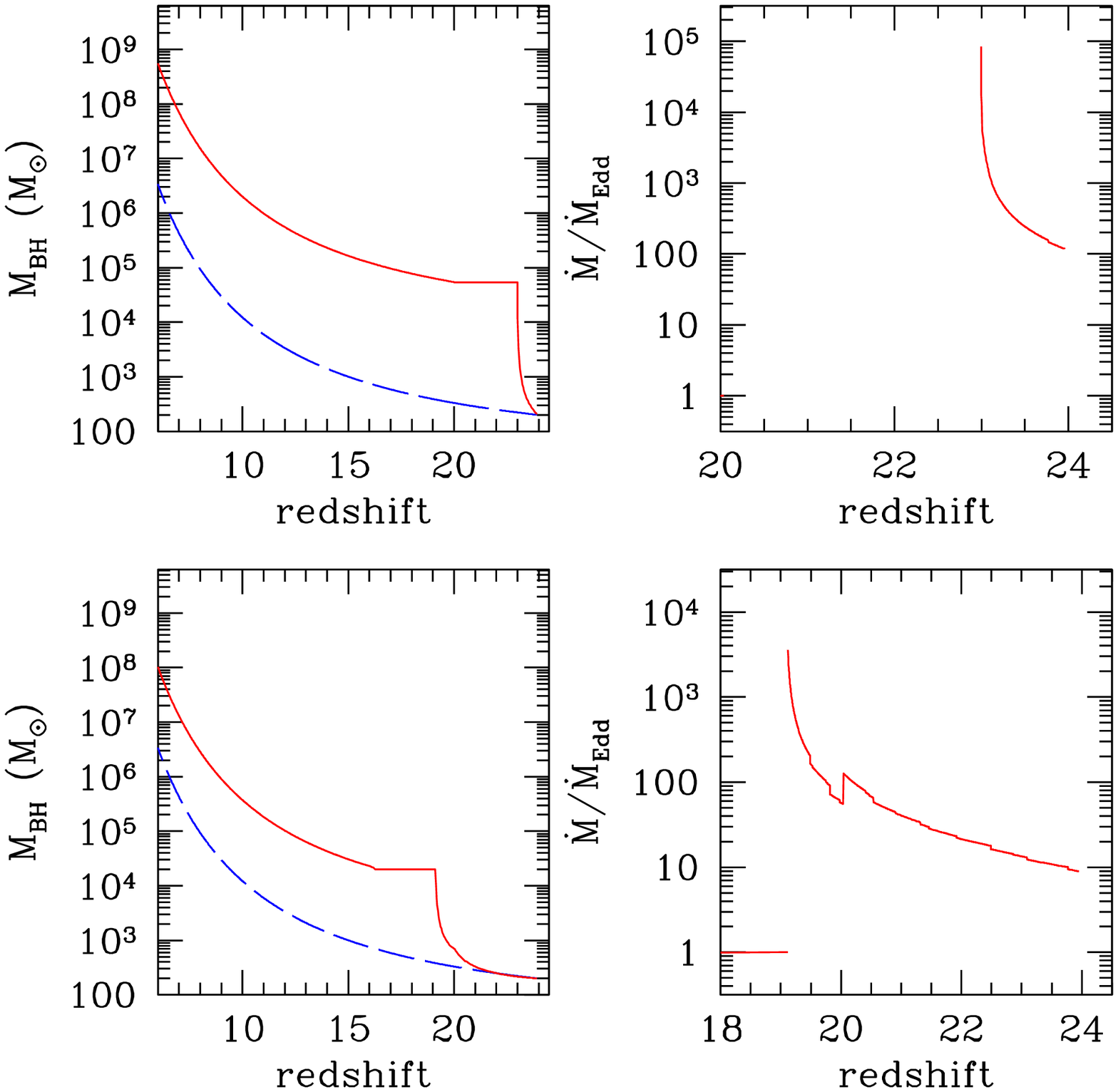}
\caption{ Evolution of the accretion rate and MBH mass growth in the main halo of the 
merger trees, $M_h=10^{13}\msun$ at $z=6$.
{\it Top:} $f_d=0.5$.
{\it  Bottom:} $f_d=0.1$.
{\it Left panel:} MBH mass as a function of redshift for the model discussed in this paper 
({\it solid line}) and assuming Eddington accretion rate at all times, with $\epsilon=0.15$
({\it dashed line}). 
{\it Right panel:} MBH accretion rate, in units of the Eddington rate.
}
\label{fig1}
\end{figure}

When the halo virial temperature  becomes larger than $10^4{\rm K}$, 
the MBH is surrounded by a coldish ($T_\gas=8000{\rm K}$) disc. The accretion rate 
is set at the Bondi rate, as determined by the gas density at the accretion radius. 
The luminosity, though, does not exceed the Eddington luminosity.
During the accretion process, the radius of the inner disc increases steeply with the 
hole mass, thus making super-Eddington accretion less likely to be sustained. We assume here 
that, when the radius of the accretion disc becomes a factor of 5 larger (though the choice 
of the exact value is somehow arbitrary) an outflow develops, blowing away the disc. 
A subsequent major merger then has to be awaited before a fresh supply of fuel 
is available to the MBH. Typically, the conditions for super-critical accretion are satisfied 
only once per MBH. If they are not, the MBH accretes at the Eddington value during the susequent
accretion episode(s). 
The radiative efficiency then evolves with the MBH spin, 
adopting the standard definition for circular equatorial orbits around a Kerr hole.
The MBHs spin is modified during the accretion phase as described in Volonteri \etal 2004.
Figure \ref{fig1} shows the evolution of MBH (mass and accretion rate) and disc in the 
main halo of one of the merger trees. This halo represents a 5-$\sigma$ density fluctuation at 
$z=24$. 
The virial temperature of the main halo is larger 
than $10^4{\rm K}$ at $z\approx 24$ and the MBH starts growing very early.  The accretion rate 
is initially super-critical by a factor of 10 and grows up to a factor of about $10^4$, thus 
making the flow more and more spherical (see Equation \ref{r_tr}). On the other hand, the 
whole 'plump' accretion disc grows in size until it crosses the trapping surface and reaches 
the assumed threshold for the end of the activity. 
The figure shows also the effect of a larger or smaller fraction of the baryons
ending up in the fat disc. A large supply of fuel ($f_d=0.5$) triggers rapid accretion early
on, but also the accretion is quenched sooner. In both cases the spin parameter is set
as $\lambda=0.05$. MBHs in halos with small spin parameter have an early and short super-critical 
accretion episode, (cfr. eq. \ref{central_density} and \ref{r_inner}) while in case of high spin 
parameter, the super-critical phase happens, typically with a more dramatic growth of the BH mass, 
at lower redshift.

\begin{figure}
\plotone{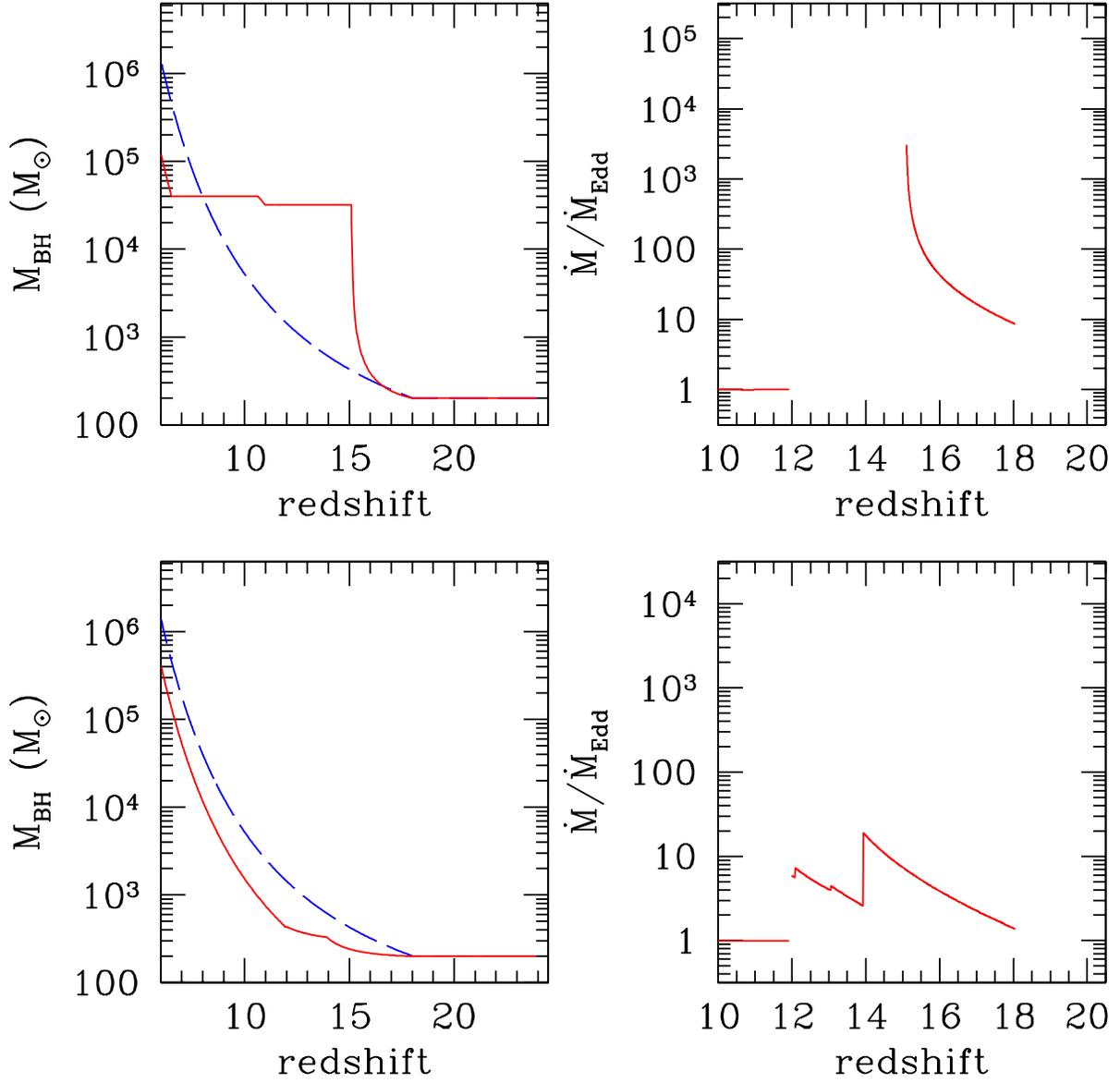}
\caption{ Same as Figure \ref{fig1}, but for a massive satellite halo. This halo is 
a 4-$\sigma$ peak in the density fluctuations field, and represents therefore the minimum mass of a 
halo which is granted a seed MBH in our model.}
\label{fig2}
\end{figure}

Our assumed threshold for MBH formation (density peaks above 4-$\sigma$) selects only the 
most massive and rare halos at a given redshift. This ensures that super-accreting systems are not 
widespread. Let us consider comparatively smaller halos (e.g. massive satellites of the main halo, 
Figure \ref{fig2}), with virial temperature below $10^4{\rm K}$ at $z\simeq24$. They accrete mass
along the merger hierarchy, until their mass implies a virial temperature larger than $10^4{\rm K}$. 
Atomic cooling, therefore, becomes effective at lower redshift, so that there is only a short 
time (if any) for the rapid growth to take place. The effect of supercriticality on black holes 
hosted in smaller halos is thus mild. Note that we have created a whole set of merger trees, and they 
give qualitatively very similar results. 

\section{Discussion}
In this paper, we have envisaged an early stage of super-critical accretion during the 
global evolution of a SMBH (see also Kauffmann \& Haehnelt 2000). 
Fuel is supplied by a dense gaseous disc forming in halos  
with $T_{\rm vir} >10^4$K, where hydrogen atomic cooling is effective.

If the disc rotates as a rigid body, the gas infalling on the central MBH
has small transverse velocity. A tiny accretion disc can form within the radius
at which radiation is trapped and the MBH can accept most of the infalling mass. 
Even if this phase lasts only until the universe in enriched by metals, the quick start 
allows the holes in the most massive halos to reach the high SMBH masses suggested by the 
SDSS quasars. It is worth noting that super-critical accretion in the fashion described
here occurs at very high redshift, and that stops well before $z\simeq 6$. We do not expect, 
therefore, that SDSS quasars are accreting above the Eddington value, consistently with
observations (Barth et al. 2003, Willott, McLure \& Jarvis 2003). A signature of 
super-critical accretion is the occurrence of outflows, that we envisage would quench activity. 
Outflows from these sources could leave their imprint by spreading metals into the IGM early-on.

One might be concerned that this model overproduces large SMBHs at low redshift 
This is not the case, since only a tiny fraction of halos have $T_\vir\gta 10^4\,$K,
when the Universe is still metal free. Assuming that metal pollution starts affecting the disc
cooling and fragmentation at $z\lta10$, only a fraction $\approx 2\times 10^{-3}$
of halos at this time is likely to be the outcome of a merger hierarchy which 
involves a seed MBH  at $z\gta20$. 
So, even if all these halos with mass $M_h\sim10^{10}\msun$ hosted a MBH with 
mass $\sim 10^6\msun$, the density
$\Omega_{\rm MBH}=(0.002\,\Omega_M\,\langle M_{\rm BH}\rangle)/{M_h}=10^{-8}$
would still be much lower than the local one.
We will discuss this issue in detail, along with the possibility that black holes can be 
displaced from galaxy centers and ejected into the IGM by the `gravitational rocket' effect 
in a subsequent paper.

As a side result, this model allows a very 'economic' re-ionization in terms of
BH seeds density. Madau \etal (2004) found that pregalactic quasars powered by MBHs 
forming in 3.5$\sigma$ peaks will reionize the IGM if they accrete at the 
Eddington rate a mass of order $10^{-3}$ the mass of the host halo in every 
major merger. Alternatively, seed holes must be more numerous (e.g. 3--$\sigma$ density fluctuations) 
at the start in order to sustain the early production of ionizing radiation. 
This latter assumption would imply that all galaxies in the local Universe with total 
mass $M_h>5\times10^{9}\msun$ are expected to host a central MBH. Allowing for a super-critical 
accretion phase, the number of ionizing photons per hydrogen atom produced by pregalactic quasars 
approaches unity at $z\gtrsim 15$, even assuming a very low initial density in seeds
(e.g. 4-$\sigma$ peaks), thus relieving the need for a very large number of seeds - 
difficult to reconcile with low redshift constraints.

The growth of SMBHs that can power SDSS quasars can be explained within a $\Lambda$CDM 
universe also assuming a more optimistic view in terms of accretion and merging.
Yoo \&  Miralda-Escud\'e (2004) showed $z\simeq 6$ quasars can be explained
assuming continued Eddington-limited accretion onto MBHs forming in halos with
$T_{\rm vir} >2000$K at $z\leq 40$. Their model assumes, also, a much 
higher influence of BH mergers in increasing the MBH mass: a contribution by itself 
of order $10^9\msun$. Their investigation takes into account the negative feedback
that dynamical processes at BH mergers (`gravitational rocket', see also Haiman 2004) 
impose on SMBHs growth. A heavy influence of MBH mergers in building-up SMBHs can
be probed by their gravitational waves emission, by the planned {\it Laser Interferometer 
Space Antenna} ({\it LISA}).

As pointed out by Shapiro (2005), explaining the presence of SMBHs at $z\simeq 6$ is 
extremely difficult if accretion occurs at the Eddington rate and via a standard thin 
disc. If alignment between the accretion disc and the MBH spin is efficient on very short 
timescales, MBHs quickly end-up maximally spinning, with a radiative efficiency 
$\epsilon\simeq 0.4$ (cfr. Equation 1). Shapiro suggests therefore that accretion occurs via 
discs where viscosity is due to magneto-hydrodynamical (MHD) turbulence. Gammie et al. (2004) 
simulations suggest in fact that the maximum spin MBHs can achieve by coupling with MHD discs is 
smaller than 1 and the corresponding maximum radiative efficiency is 
$\epsilon\simeq 0.19$. The gain in time available for SMBH growth is less than a factor of 3, 
though, thus probably requiring that dynamical negative feedback is not important. 

The issues we have discussed in this paper will be clarified by better evidence 
on mini-quasars and on the re-ionization history at $z\simeq 10$. Future missions 
such as {\it LOw Frequency ARray} ({\it LOFAR}) and the {\it Square Kilometer Array} 
({\it SKA}) will probe directly the IGM up to high redshift. 
Our discussion also provides added motivation for ongoing studies of the flow patterns that occur 
when gas falls towards a hole at a supercritical rate.

{}


\end{document}